\documentclass[11pt, oneside]{amsart}
\usepackage[utf8]{inputenc}
\usepackage[left=0.8in, right=0.8in, top=0.8in, bottom=0.9in, centering]{geometry}      
\usepackage{float, epsfig, natbib, lineno, appendix}
\usepackage{color, amssymb, amsmath, amsthm, verbatim, wasysym, mathrsfs}
\usepackage[mathscr]{eucal}
\usepackage{hyperref} 
\usepackage[font={footnotesize}]{caption}
\usepackage[dvipsnames]{xcolor}
\usepackage{scalerel, stackengine}
\setstackEOL{\#}
\stackMath
\def\hatgap{2pt}
\def\subdown{-2pt}
\newcommand\reallywidehat[2][]{ \renewcommand\stackalignment{l} \stackon[\hatgap]{#2}{ \stretchto{
    \scalerel*[\widthof{$#2$}]{\kern-.6pt\bigwedge\kern-.6pt}
    {\rule[-\textheight/2]{1ex}{\textheight}}}
    {0.5ex}_{\smash{ \belowbaseline[\subdown]{\scriptstyle#1} }}
}}

\newcommand{\com}		{\, ,}
\newcommand{\per}		{\, .}
\newcommand{\defn}	    {\ensuremath{\stackrel{\mathrm{def}}{=}}}
\newcommand{\beq}		{\begin{equation}}
\newcommand{\eeq}		{\end{equation}}

\newcommand{\mystrut}[1]{\vrule width0pt height0pt depth#1\relax}

\renewcommand{\d}		{\partial}

\newcommand{\lap}		{\triangle}

\newcommand{\ii}	    {\mathrm{i}}
\newcommand{\cc}	    {\star}
\newcommand{\dd}	    {{\rm d}}
\newcommand{\id}	    {{\, \rm d}}

\newcommand{\at}[1]     {\, |_{#1}}

\begin{document}

\title{ \vspace{-6ex}
Deep flows transmitted by forced surface gravity waves
\vspace{-0.6ex} }

\author{Nick Pizzo}
\author{Gregory L Wagner* }
\thanks{* The authors contributed equally to this work.}

\maketitle
\section*{Abstract}

We examine a \textcolor{black}{two-dimensional} deep-water surface gravity wave packet generated by a pressure disturbance in the Lagrangian reference frame. The pressure disturbance has the form of a narrow-banded weakly nonlinear deep-water wave packet. During forcing, the vorticity equation implies that the momentum resides entirely in the near-surface Lagrangian-mean flow, which in this context is often called the ``Stokes drift''. After the forcing turns off, the wave packet propagates away from the forcing region, carrying with it most of the energy imparted by the forcing. These waves together with their induced long wave response have no momentum \textcolor{black}{in a depth integrated sense}, in agreement with the classical results of \citet{LH1962} and \citet{Mcintyre1981}. The total flow associated with the propagating packet has no net momentum, in agreement with the classical results of \citet{LH1962} and \citet{Mcintyre1981}. In contrast with the finite-depth scenario discussed by \citet{Mcintyre1981}, however, momentum imparted to the fluid during forcing resides in a dipolar structure that persists in the forcing region --- rather than being carried away by shallow water waves. We conclude by examining waves propagating from deep to shallow water and show that wave packets, which initially have no momentum, may have non-zero momentum in finite-depth water through reflected and trapped long waves. This explains how deep water waves acquire momentum as they approach shore. \textcolor{black}{The artificial form of the parameterized forcing from the wind facilitates the thought experiments considered in this paper, as opposed to striving to model more realistic wind forcing scenarios.}

\section{Introduction}

Much of the ocean's momentum is transferred from the atmosphere by form stresses pressing on the faces of surface gravity waves \citep{Grare2013, Melville1996}.
Yet the fluid dynamical details of how atmospheric pressure is transmitted across the surface --- somehow simultaneously forcing a current beneath the surface while also exciting a surface gravity wave --- are mysterious \citep{Pizzo2021}.
In this paper we elucidate some of these details through a detailed asymptotic analysis of the forced growth of a \textcolor{black}{two-dimensional} surface gravity wave packet of finite extent by an idealized atmospheric pressure distribution.

The unforced cousin of this problem was analyzed by \citet{LH1962}, who computed the total second-order mean flow associated with a freely-propagating deep-water surface gravity wave packet.
\citet{LH1962} showed that the surface-wave-induced mean flow has two components: first, a shallow surface component with no Eulerian-mean signature wholly attributable to the surface gravity wave Stokes drift, which may be evaluated diagnostically from and has similar vertical scales as the primary wave field.
The second component of the mean flow is driven by the convergence and divergence of the primary flow, is purely Eulerian-mean, and is longer and deeper with scales similar to that of the wave \textit{packet}.
Remarkably and as emphasized by \cite{Mcintyre1981}, the total vertically-integrated momentum of the freely-propagating packet vanishes at every horizontal point.

The absence of momentum in freely-propagating wave packets begs the question: where is the momentum transferred from the atmosphere during wave generation?
\citet{Mcintyre1981} argues that the momentum transferred during wave generation is carried away by packet-scale, shallow surface gravity waves --- which being shallow, \textit{can} carry momentum.
But \citet{Mcintyre1981}'s answer, in addition to lacking the details that might accompany a more systematic solution of the equations of motion, produces counter-intuitive outcomes if the depth of the fluid is increased while holding the scale of the packet and atmospheric forcing fixed.
For example, in \cite{Mcintyre1981}'s theory atmospheric momentum would have to be instantaneously transferred farther and farther from the disturbance, carried by shallow-water waves with longer and longer wavelengths without bound as the depth of the fluid increases.

This paper revisits \textcolor{black}{the momentum} question through a detailed and systematic asymptotic solution of the Lagrangian equations of motion.
Our aim is to elucidate the fate of atmospheric momentum transferred to a fluid during surface gravity wave packet generation, and in process, clarify \cite{Mcintyre1981}'s argument.
We find, contrary to \cite{Mcintyre1981}, that in an infinitely deep, non-rotating, and unstratified fluid, atmospheric momentum is not transferred to propagating waves but instead remains \textit{in place} after forcing ceases while the packet propagates away.
Moreover, we find the precise mechanism by which this occurs: as the packet grows, the pressure forces that drive its deep return flow are opposed --- pointwise within the interior of the fluid --- by the packet-scale component of the atmospheric pressure gradient.
As a result, there is no deep return flow \textit{while} the packet is forced, and it is only at the moment the forcing ceases that the deep return flow and its counterpart --- the part that contains the net momentum transferred from the atmosphere --- remains behind.
We call this part of the flow, which is effectively transmitted from the surface into the deep ocean during the generation of surface gravity waves, and is given precisely as the negative of the deep return flow associated with the propagating packet at the moment the forcing ceases, the ``wave-transmitted flow''.

Our results are obtained through a multi-scale (in time and depth) asymptotic expansion of the equations of fluid motion formulated in a Lagrangian reference frame, which differs from usual analyses that leverage the Eulerian reference frame \citep{LH1962, van2016, Haney2017}.
In our Lagrangian analysis, the vorticity constraint --- in our case, the vorticity of the fluid must vanish --- plays a prominent role in identifying the shallow and deep components of the mean response to the atmospheric forcing \citep{Salmon2020, Pizzo2023}.
This analysis highlights the close connection between the mean Lagrangian motion, the momentum of the fluid, and the vorticity of the fluid.
As these flows are irrotational it might seem surprising that a vorticity analysis clarifies the dynamics -- but indeed this perspective has proven to be useful in several contexts for water waves \citep{Pizzo2023, Blaser2024} because in two dimensions vorticity is conserved on fluid particles, and so it is closely related to the mass flux.


An additional point of confusion arises from a close reading of \citet{LH1962} and \citet{Mcintyre1981}.
Our analysis shows that when both the primary wave packet and the long wave response are in deep-water, neither carries momentum. 
Yet, shallow-water waves \textit{can} carry momentum, which seemingly leads to a violation of conservation of momentum if the wave packet propagates from deep to shallow-water.
To better understand this, we consider a wave packet in infinitely deep-water which experiences a step to finite depth and argue that this leads to the long wave being partially trapped at the step. When the momentum in each component is accounted for, the total momentum of the system remains conserved. 

\begin{figure}
    \centering
    \includegraphics[width = 0.5\textwidth]{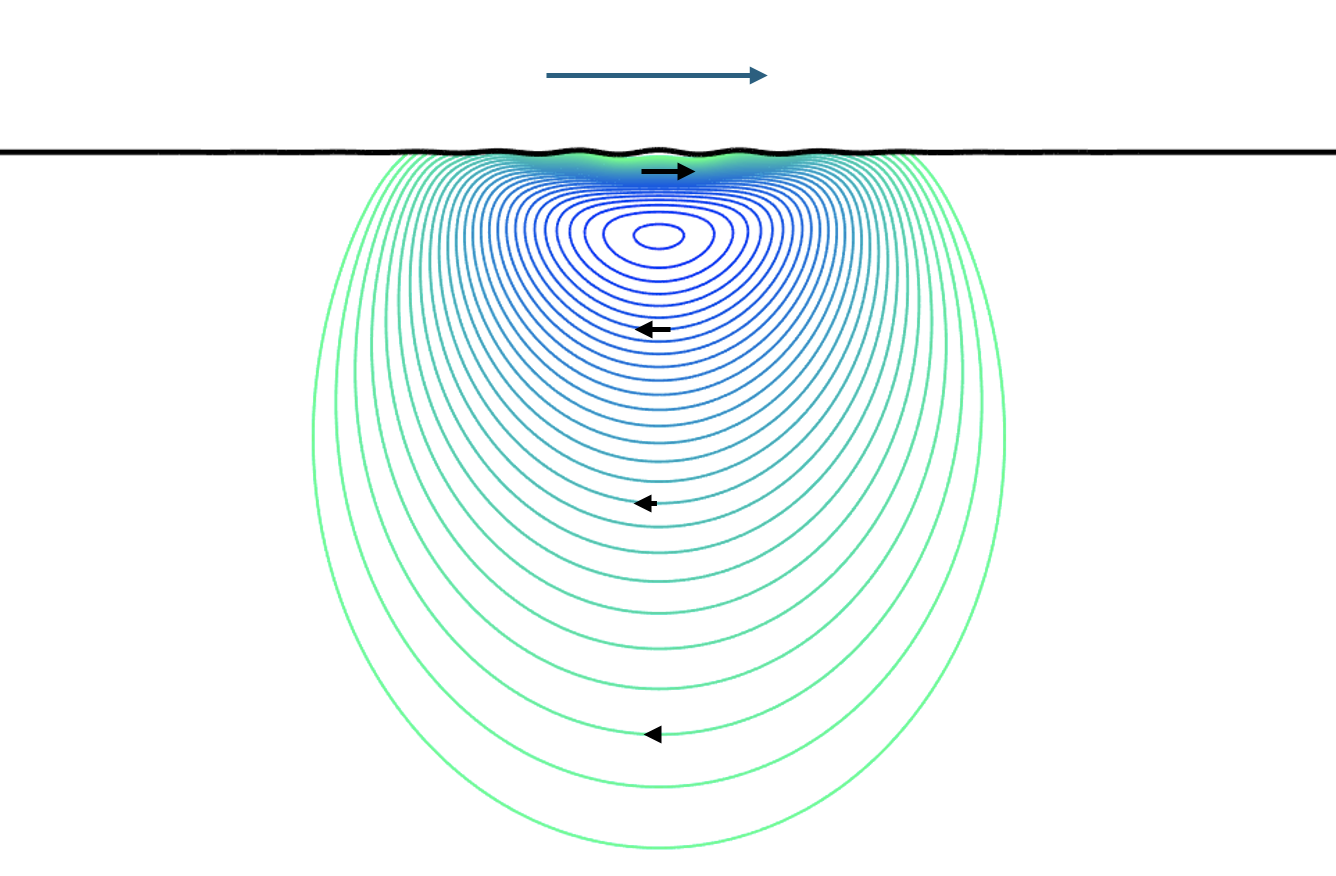}
    \caption{Contours of the streamlines in the Lagrangian reference frame for the second order mean flow of a deep-water surface gravity wave packet travelling from left to right, as indicated by the arrow above the wave packet. The colorbar indicates the value of the streamline contours, with blue being larger value than green. The velocity is sketched by the black arrows. The broad long wave response to the wave packet decays much more slowly with depth than the response to the carrier waves and is such that the momentum in the wave packet is exactly zero. }
    \label{fig:experiment}
\end{figure}

The outline of this paper is as follows: in \S 2 we present the equations of motion and several conservation laws for deep-water weakly nonlinear narrow banded surface gravity waves in the Lagrangian reference frame.
In \S 3, we satisfy the equations of motion to third order in a small parameter.
\S 4 considers the evolution of the waves and the mean flow after the forcing is turned off.
The conclusions are presented in \S 5. 

\section{Equations of motion}

We consider two-dimensional inviscid deep-water surface gravity waves. The Euler equations in Lagrangian coordinates are \citep[][\S 13]{Lamb1932}
\begin{equation}\label{x_mom}
\mathcal{J}x_{\tau\tau} + p_a y_b - p_b y_a = 0 \com
\end{equation}
and
\begin{equation}\label{y_mom}
\mathcal{J}y_{\tau \tau} + p_b x_a - p_a x_b +\mathcal{J} g=0 \com
\end{equation}
where $(x,y)$ are the particle locations, $(a,b)$ are particle labels, $p$ is the fluid pressure, $g$ is the acceleration due to gravity, and the specific volume $\mathcal{J}$ is defined as
\begin{equation}\label{mass}
\mathcal{J} \defn x_ay_b-x_by_a.
\end{equation}
Derivatives with respect to the time $\tau$ take $(a,b)$ to be fixed, so that $\partial_{\tau}\defn D/Dt$. Even though the acceleration term is linear in the Lagrangian frame, the pressure gradient terms in \eqref{x_mom}--\eqref{y_mom} are now nonlinear. This highlights the fact that the mapping between the two frames may be (highly) nonlinear. 

The continuity equation in Lagrangian coordinates is
\begin{align}
0 &= u_x + v_y \com \\
& = \frac{\partial (x_{\tau},y)}{\partial(x,y)}-\frac{\partial (y_{\tau},x)}{\partial(x,y)} \com \\
&= \left(\frac{\partial (x_{\tau},y)}{\partial(a,b)}+\frac{\partial (x,y_{\tau})}{\partial(a,b)}\right)\frac{\partial(a,b)}{\partial(x,y)} \com
\end{align}
which implies 
\beq
\mathcal{J}_\tau = 0 \per
\eeq
In order for the mapping to be invertible we require that $\mathcal{J}$ does not change sign. However, there are no other restrictions on $\mathcal{J}$, and this gauge invariance to how one labels the particles offers tremendously algebraic simplifications if the correct gauge is chosen. This symmetry also corresponds to the conservation of vorticity, $q$ \citep{Salmon1988}.

The vorticity $q$ is given by
\begin{align}
q &\defn v_x - u_y \com \\
&= \frac{\partial (v,y)}{\partial(x,y)}+\frac{\partial (u,y)}{\partial(x,y)} \com
\end{align}
so that
\begin{equation}
q\mathcal{J}=\frac{\partial(x_{\tau},x)}{\partial(a,b)}+\frac{\partial(y_{\tau},y)}{\partial(a,b)},
\end{equation}
where the Euler equations in two-dimensions imply that
\beq
q_\tau = 0 \per
\eeq
For irrotational flows, $q=0$. 

In the absence of forcing, we require that the pressure $p$ vanish at the free surface, i.e. $p=0$ at $b=0$. Finally, we require that $y_{\tau}\to 0$ as $b\to -\infty$ so that flow vanishes at depth. A word of warning concerning nomenclature -- we say waves are deep when $kd\gg 1$ for $k$ their wavenumber and $d$ the depth of the fluid while we say a wave packet is in deep-water if $\Delta k d \gg 1$ for $\Delta$ the non-dimensional bandwidth of the wave packet, here taken to be $\epsilon$. There can be scenarios where the waves themselves are in deep water, but the wave packets are not.

The kinematic boundary condition is automatically satisfied in the Lagrangian reference frame. Although not pursued in detail here, the series convergence rates are faster in the Lagrangian frame \citep{LHC1976, Clamond2007} than the Eulerian frame as particles tend to cluster in the crests. 

\subsection{Conservation laws}

Vertically integrating equation (\ref{x_mom}) yields
\begin{equation}\label{mom_int}
\partial_\tau \int_{-\infty}^0 \mathcal{J}x_{\tau}  \dd b +\int_{-\infty}^0\left( p_ay_b-p_by_a\right)  \dd b = 0. 
\end{equation}
We rewrite the second term in \eqref{mom_int} by noting
\begin{equation}
\frac{1}{2} \partial_a \int_{-\infty}^0 \left(py_b-p_by \right)  \dd b=\frac{1}{2} \int_{-\infty}^0\left(p_ay_b-p_by_a +py_{ab}-p_{ab}y \right)  \dd b.
\end{equation}
Further integrating by parts produces
\begin{equation}
 \frac{1}{2} \partial_a \int_{-\infty}^0 \left(py_b-p_by\right)   \dd b= \int_{-\infty}^0 \left(p_ay_b-p_by_a\right)   \dd b + \frac{1}{2}\left(py_a-yp_a)\right|_{b=0},
\end{equation}
so that the $b$-integrated conservation of horizontal momentum expressed by equation~\eqref{mom_int} becomes 
\begin{equation}
\partial_\tau \int_{-\infty}^0\mathcal{J} x_{\tau} \ \dd b+\partial_a \int_{-\infty}^0 py_b  \dd b= \frac{1}{2}\left(yp_a-py_a)\right|_{b=0}.
\label{mom_cons}
\end{equation}
In the absence of pressure forcing at the surface, $p=p_a=0$ at the free surface so that the right hand side becomes zero. 
The second term here represents the $x$-momentum flux in this framework, which we define as $S$, where
\begin{equation}
S \defn \int_{-\infty}^0 py_b\  \dd b,
\label{mom_flux}
\end{equation}
so that 
\begin{equation}
\mathcal{I}_\tau + S_a = \frac{1}{2}\left(y p_a - p y_a) \right|_{b=0},
\end{equation}
with the impulse $\mathcal{I}$ defined as
\beq
\mathcal{I} \defn \int_{-\infty}^0 \mathcal{J}x_{\tau} \dd b.
\eeq
In the absence of pressure forcing, we have 
\beq
\mathcal{I}_\tau + S_a = 0
\eeq
so that the momentum is conserved. 

We derive an energy integral by forming $x_\tau \eqref{x_mom} + y_\tau \eqref{y_mom}$ and vertically integrating to find
\beq
0 = \partial_\tau \int_{-\infty}^0\mathcal{J}
    \left( \frac{x_{\tau}^2 + y_{\tau}^2}{2} + y \right ) \dd b
    + \int_{-\infty}^0 \left[ p_a \left( x_{\tau} y_b -y_{\tau}x_b\right)+p_b(y_{\tau}x_a-x_{\tau}y_a) \right ] \, \dd b \per
\eeq
Now, we begin by considering the integral 
\beq
\partial_a \int_{-\infty}^0 p_b(xy_{\tau}-x_{\tau}y) +p(xy_{\tau b}-x_{\tau b}y) \dd b \per
\eeq
We may then show, by application of integration by parts that 
\beq
\begin{split}
& \int_{-\infty}^0 \left [ p_a(x_{\tau}y_b-y_{\tau}x_b)+p_b(y_{\tau}x_a-x_{\tau}y_a) \right ] \, \dd b \\
& \qquad = \partial_a \int_{-\infty}^0
    \left [ p_b(xy_{\tau}-x_{\tau}y) +p(xy_{\tau b}-x_{\tau b}y) \right ] \, \dd b
    +\left [ p_a \left (xy_{\tau}-x_{\tau}y \right ) \right]_{b=0},
\end{split}
\eeq
so that  
\begin{equation}
E_\tau + F_a = \left[ p_a \left ( x_{\tau} y - x y_{\tau} \right ) \right ]_{b=0},
\end{equation}
where the energy is 
\begin{equation}\label{en_int}
E \defn \int_{-\infty}^0\mathcal{J} \left(\frac{x_{\tau}^2+y_{\tau}^2}{2}+y\right)\dd b,
\end{equation}
and the energy flux $F$ is given by 
\begin{equation}
F \defn \int_{-\infty}^0 p(x_by_{\tau}-x_{\tau}y_b) \dd b.
\end{equation}
In the absence of pressure forcing the energy is conserved.

Next, consider the kinematic relationship $z_{\tau}^{\cc}=\chi_z$, for $\cc$ denoting complex conjugate and $\chi =\phi+\ii \psi$ the complex velocity potential. If we then map to the Lagrangian frame, and let $s=a+\ii b$, we have, $z_{\tau}^{\cc}z_s = \chi_s$ \citep{LH1980} so that the stream function $\psi$ in the Lagrangian reference frame satisfies 
\beq
\psi_a = x_{\tau}y_a-y_{\tau}x_a; \quad \psi_b = x_{\tau}y_b-y_{\tau}x_b.
\eeq
From this association, we can then write the energy flux as
\beq
F= \int_{-\infty}^0 p \, \psi_b \, \dd b.
\eeq

\subsection{Momentum and energy transfer by surface pressure forcing}

We now consider the total momentum conservation by integrating \eqref{mom_cons} over $a$ and assuming that the pressure is compact so that upon integrating by parts we find
\beq \label{18}
\partial_\tau \iint \mathcal{J} u \id a \id b = \int p y_a \at{b=0} \id a \per
\eeq
The left side of \eqref{18} is the rate of change of total horizontal momentum. The right side of \eqref{18} is the form stress: the only way by which momentum enters an inviscid fluid in a closed system.

Similarly, we can look at the dynamical budget. Once again, we average over all horizontal particle labels to find 
\beq
\partial_\tau \iint  \mathcal{J} \left(\frac{x_{\tau}^2+y_{\tau}^2}{2}+y\right) \id a \id b =\int \left.p(x_{\tau}y_{a}-x_{a}y_{\tau})\right|_{b=0} \, \dd a \per
\eeq
The right hand side of the momentum balance is a term related to the form drag, while the right hand side of the energy equation is related to the spin and angular momentum of the fluid \citep{LH1980b}.

\section{Weakly nonlinear narrow-banded waves}

Our goal is to find $(x,y,p)$ to third order in the small parameter $\epsilon \ll 1$. First, we satisfy conservation of mass and vorticity. We then find an expression for the pressure $p$. Finally, the dispersion relationship is found by ensuring that the pressure condition is met at $b=0$. 

To solve our system we expand $x,y,p$ in asymptotic series,
\beq
x=\sum_{n} \epsilon^n x^{(n)}; \quad 
y=\sum_{n} \epsilon^n y^{(n)}; \quad 
p=\sum_{n} \epsilon^n p^{(n)},
\eeq
where $x^{(n)} =x^{(n)}(a,b,\tau,\epsilon a, \epsilon b, \epsilon \tau,...), y^{(n)}=y^{(n)}(a,b,\tau,\epsilon a, \epsilon b, \epsilon \tau,...),p^{(n)}=p^{(n)}(a,b,\tau,\epsilon a, \epsilon b, \epsilon \tau,...)$. Here, the superscript with parentheses represents the $n-th$ term in the expansion, not a power of that variable.   

In our analysis we take $\mathcal{J}=1$, which amounts to a gauge choice \citep{Salmon2020} sometimes known as Miche's condition \citep{Clamond2007}. 
The pressure forcing $p^\mathrm{atm} $ is defined at the free surface, $b=0$, at first order to be 
\beq
p^\mathrm{atm} = 2g\epsilon  \mathcal{A}\sin \theta ,
\eeq
where the phase $\theta$ is defined
\beq 
\theta \defn k a - \underbrace{\mystrut{1.2ex} \sqrt{g k} }_{\defn \omega} \tau \com
\eeq
and $\mathcal{A}=\mathcal{A}(\epsilon a, \epsilon \tau)$ is a slowly varying function (with units of length), which represents the modulation to the forcing event in space and time.  
For example, figure~\ref{fig:experiment} is made with $\mathcal{A} = \exp \left \{ - \left ( a - c_g t \right )^2 / 2 \delta^2 \right \}$, where $c_g$ is the group velocity and the packet width $\delta$ is $O(\epsilon^{-1})$.

\subsection{Zeroth order expansion}
At zeroth order, the system is in hydrostatic balance and we have 
\beq
x^{(0)} = a; \quad y^{(0)} = b; \quad p^{(0)} =-gb.
\eeq
\subsection{First order expansion}

To first order in $\epsilon$, conservation of mass requires
\beq
x^{(1)}_{a}+y^{(1)}_{b}=0,
\label{mass-1}
\eeq
while the vorticity equation implies 
\beq
-x^{(1)}_{\tau b}+y^{(1)}_{\tau a}=0.
\eeq
That is, the first order fluid velocity $(x^{(1)}_{\tau},y^{(1)}_{\tau})$ satisfies the Cauchy-Riemann equations, with each component being harmonic with respect to $(a,b)$. 
 
The $O(\epsilon)$ momentum equations are
\begin{gather}
    \label{a-momentum-1}
    u^{(1)}_{ \tau} + p^{(1)}_{a} + g x^{(1)}_{b} = 0 \com \\
    \label{b-momentum-1}
    v^{(1)}_{\tau} + p^{(1)}_{b} + g y^{(1)}_{b} = 0 \per
\end{gather}

Forming $\d_a \eqref{a-momentum-1} + \d_b \eqref{b-momentum-1}$ and using \eqref{mass-1} yields
\beq \label{harmonic-pressure-1}
\lap p^{(1)} = 0 \per
\eeq
\eqref{harmonic-pressure-1} has the surface boundary condition $p^{(1)} \at{b=0} = p^\mathrm{atm} = 2 g \mathcal{A} \sin \theta$.
The fluid has infinite depth so that we apply the condition that $p^{(1)} \at{b \to -\infty} \to 0$.
The solution that satisfies \eqref{harmonic-pressure-1}, $p^{(1)} = 2 g \mathcal{A}\sin \theta$ at $b=0$ and $p^{(1)} \to 0$ as $b \to -\infty$ is
\beq
p^{(1)} = 2g \mathcal{A}\sin \theta e^{kb} 
\eeq
where we have assumed that $k > 0$, that is all waves are travelling to the right in our system.

We differentiate the momentum equations once more with respect to time, and use the Cauchy-Riemann equations to find
 \beq
 u^{(1)}_{\tau \tau}+gu^{(1)}_{b}=-p^{(1)}_{\tau a},
 \eeq
 \beq
 v^{(1)}_{\tau \tau}+gv^{(1)}_{b}=-p^{(1)}_{\tau b}.
 \eeq
Taking $(u^{(1)},v^{(1)})$ to be separable with vertical dependence $e^{kb}$ implies 
 \beq
 u^{(1)}_{\tau \tau}+gku^{(1)}=-p^{(1)}_{\tau a},
 \eeq
 \beq
 v^{(1)}_{\tau \tau}+gkv^{(1)}=-p^{(1)}_{\tau b}.
 \eeq 
This is the governing equation for a (resonantly) forced harmonic oscillator. Taking the initial conditions to be that the surface is initially flat, i.e. $(x^{(1)},y^{(1)})=(0,0)$ at $\tau=0$, this has solution
\beq
x^{(1)} = -\frac{\mathcal{A}\omega \tau}{2}\cos \theta e^{kb}
\eeq
\beq
y^{(1)} = \frac{\mathcal{A}\omega \tau}{2}\sin \theta e^{kb}. 
\eeq
The free surface is growing secularly in time to first order in $\epsilon$. By forcing the system with a wave form that exactly solves the equations of motion at this order, we eliminate the need to resolve dispersive corrections (c.f. the Cauchy-Poisson problem) that would unnecessarily complicate the analysis. 

\subsection{Second order expansion}

To second order in $\epsilon$, we find that mass conservation implies 
\beq
x^{(2)}_{a}+y^{(2)}_{b}=\omega^2 \tau^2k^2 \mathcal{A}^2e^{2kb}
\label{sec_mass}
\eeq
and vorticity conservation implies 
\beq
x^{(2)}_{\tau b}-y^{(2)}_{\tau a}=2\omega \tau^2k^2 \mathcal{A}^2 e^{2kb}.
\eeq
By differentiating \eqref{sec_mass} with respect to time, these equations can be reduced to 
\beq
\Delta u^{(2)} =4\omega^3 \tau^2 k^3 \mathcal{A}^2 e^{2kb},
\label{u2}
\eeq
\beq
\Delta v^{(2)} =4 \omega^2 \tau k^3 \mathcal{A}^2 e^{2kb}.
\eeq
Unlike in the Eulerian frame, at second order these equations have no wavy terms (in other words, no dependence on the rapidly-varying phase $\theta$) and therefore represent the \textit{mean} flow response to the pressure forcing. 

We are interested in solving for the behavior of $u^{(2)}$, so we must find a boundary condition on this term. The horizontal momentum equation implies that at second order
\beq
 u^{(2)}_{\tau } =-p^{(2)}_{a}-gy^{(2)}_{a} +2g\omega \tau k^2 \mathcal{A}^2. 
\eeq
Taking a spatial average, $(2\pi)^{-1}\int_0^{2\pi}\dd a$, at $b=0$ this reduces to 
\beq
u^{(2)} =  \omega \tau^2 k\mathcal{A}^2,
\label{bc2}
\eeq
while we also require $u^{(2)} \to 0$ as $b\to-\infty$. 

Equation \eqref{u2} with boundary condition \eqref{bc2} has solution 
\beq
u^{(2)} = \frac{\omega}{k} k^2 \mathcal{A}^2  \omega^2 \tau^2 e^{2kb}.
\eeq
The mean $x$-momentum of the system is 
\beq
\left \langle \int_{-\infty}^0 \mathcal{J}x_{\tau} \dd b\right \rangle = \epsilon^2 \int_{-\infty}^0 \mathcal{J}u^{(2)} \dd b= \epsilon^2 \frac{\omega}{2k} k^2\mathcal{A}^2 \omega^2 \tau^2  +O(\epsilon^4),
\eeq
where the angle brackets are a spatial average over one period. This integral is non-zero and implies that the momentum is entirely contained in the drift of the carrier waves. Note, the phase averaged momentum is conserved in a local sense at each horizontal location (in the slow variable $\epsilon a$), as well as in a global sense when integrating over all of $a$. To interpret these results it is instructive to juxtapose this with the unforced scenario. 

\begin{figure}
    \centering
    \includegraphics[width = 0.5\textwidth]{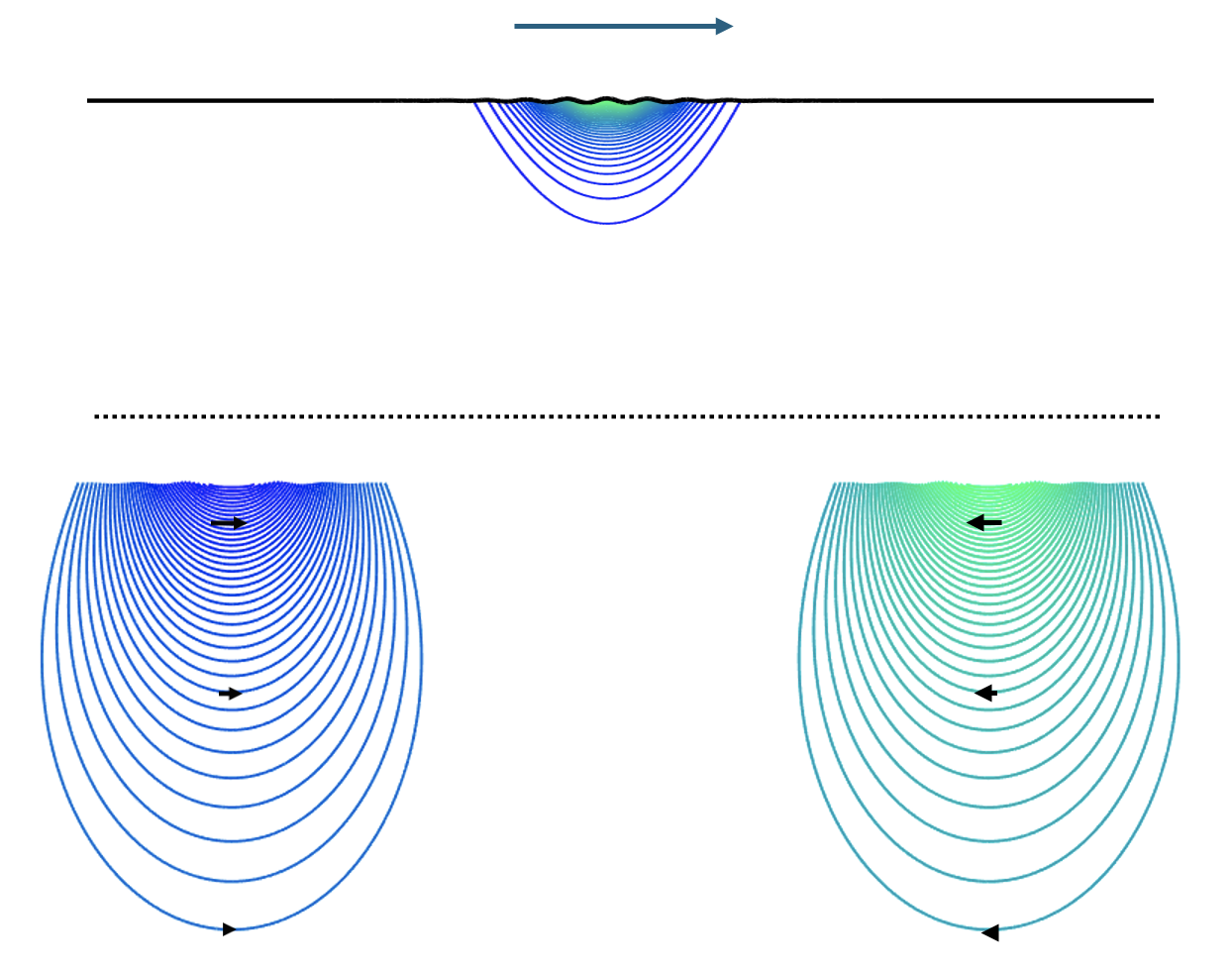}
    \caption{The streamlines for the second order mean flow of a forced deep-water surface gravity wave packet travelling from left to right, as indicated by the arrow above the wave packet. During forcing there is no long wave response to the wave packet. This is interpreted as the forcing generating an equal and opposite return flow that exactly cancels the long wave, resulting in their only being the carrier wave drift in the forcing region. This is shown below the dashed line.  }
    \label{fig:experiment}
\end{figure}
 
\section{Behavior after forcing is turned off}

We now turn off the forcing at a time $\tau=T$ and consider a time when the packet as propagate sufficiently far that we can divide the total flow into two components: a component $\psi_C$ that propagates (is ``carried'' by) with the packet, and a second component ``wave-transmitted'' component $\psi_{WT}$ that remains at the forcing site.

\subsection{The mean flow component co-propagating with the wave packet}

To compute $\psi_C$, we consider the mean flow associated with a freely-propagating packet, following \citet{Salmon2020} closely.
At first order, 
\beq
x^{(1)}=-\omega T\mathcal{A}e^{kb} \cos\theta,
\eeq
\beq
y^{(1)}=\omega T\mathcal{A}e^{kb} \sin\theta,
\eeq
\beq
p^{(1)}=0,
\eeq
and $\omega^2=gk$. These waves are no longer growing in time and represent linear deep-water surface gravity waves. 
At second order, the equations governing the mean flow become 
\beq
x^{(2)}_{a}+y^{(2)}_{b}=\omega^2 T^2k^2 \mathcal{A}^2e^{2kb}
\eeq
and vorticity conservation implies 
\beq
x^{(2)}_{\tau b}-y^{(2)}_{\tau a}=2\omega T^2k^2 \mathcal{A}^2 e^{2kb}.
\eeq
Differentiating the mass conservation equation with respect to time, we have 
\beq
x^{(2)}_{\tau a}+y^{(2)}_{\tau b}=0+O(\epsilon^3),
\eeq
which implies we can define a stream function for the second order mean flow (a linearization of equation 26):
\beq
x^{(2)}_{\tau} = -\psi_b;\quad y^{(2)}_{\tau}=\psi_a.
\eeq
The equation governing the mean flow becomes
\beq \label{wave-borne-flow}
\Delta \psi_C = -2 k^2 \omega^3 T^2 \mathcal{A}^2 e^{2kb}.
\eeq
Next, the pseudomomentum is defined as 
\beq
\bold{P}=\bold{k}\tilde{\mathcal{A}}=\bold{k}E/\omega,
\eeq
where $\tilde{\mathcal{A}}=\omega T \mathcal{A}$. Therefore, the right hand side of \eqref{wave-borne-flow} is the curl of the pseudomomentum: 
\beq
\boldsymbol{\nabla_a} \times \bold{P}=-2 k^2 \omega^3 T^2 \mathcal{A}^2 e^{2kb}\hat{z},
\eeq
where $\boldsymbol{\nabla_a} \defn (\partial_a,\partial_b)$ and $\hat{z}$ points out of the page. 

The boundary condition for $\psi$ is found once again from the horizontal momentum equation and is $\psi =0$ at $b=0$. This immediately implies that the total momentum is zero, as 
\beq
\frac{1}{2\pi}\int_0^{2\pi}\int_{-\infty}^0 \mathcal{J}x_{\tau} \dd b\dd a =\psi(b=0)+O(\epsilon^3).
\label{no_mo}
\eeq
We can consider the particular and homogeneous solution to this system. By inspection,
\beq \label{stokes-drift}
\psi_S =  -\omega^3 T^2 \mathcal{A}^2 e^{2kb},
\eeq
agreeing with the classical prediction of Stokes. Using $\psi_C = \psi_S + \psi_D$ and plugging \eqref{stokes-drift} into \eqref{wave-borne-flow} implies that $\psi_D$ obeys
\beq
\Delta \psi_D = 0
\eeq
with boundary condition $\psi_D(b=0) = -4\omega^3 T^2 \mathcal{A}^2$. The domain of the labels here is the lower half plane, so that we may use the theory of Poisson kernels to write down the solution 
\beq
\psi_D = -\frac{4\omega^3 T^2 k^2}{\pi}\int_{-\infty}^{\infty} \frac{\mathcal{A}^2 b}{(a'-a)^2+b^2}\dd a'.
\eeq
$\psi_S$ and $\psi_D$ are used to generate the figures in this manuscript. Note, some care must be taken at $b=0$, where $\psi_D$ is not zero but rather reduces to the prescribed boundary condition with $b (1+(a-a')^2/b^2)^{-1}$ being a $\delta-$function sequence as $b\to 0$.

The depth dependence of the long wave response is an integral of algebraic functions, as opposed to the exponential decay of the carrier waves. This is reflected in the fact that the contours in figure 1 near the surface have a much larger gradient with respect to label $b$ than the long wave response. In the limit that the amplitudes are uniform except at the edges of the packet, we return to the image presented in \citet{Mcintyre1981}. 

The energy is still entirely contained in wave motion -- that is the return flow does not contribute to this budget at this order. Similarly, one can show that the momentum flux, $S$ as defined by the phase average of \eqref{mom_flux}, is entirely supported by the carrier wave motion and has no contribution form the long wave response.

\subsection{The stationary forced response}

The above analysis corroborates the computations by \citet{LH1964} and \citet{Mcintyre1981} and offers a path towards interpreting the results from the forcing case, which is shown in figure 2. As the mean flow is entirely in the carrier wave drift during forcing, and the mean flow is linear at this order, we conclude that there is an equal and opposite ``wave-transmitted'' flow that exactly cancels $\psi_D$ while the forcing is active. In other words, up to the moment that the forcing ceases there exists a mean flow
\beq
\psi_F |_{t=T} = - \psi_D |_{t=T} \com
\eeq
that contains the momentum transmitted by the pressure forcing, through the wave field, and into the fluid below.

But this is not the whole story.
The pressure condition for the forced flow changes after the pressure forcing turns off, such that $p=0$ at $b=0$.
As a result, $\psi_F$ does not satisfy the equations of motion on its own --- the total wave-transmitted flow $\psi_{WT}$ that remains at the forcing site must include a fourth component such that
\beq
\psi_{WT} = \psi_F + \psi_4 \per
\eeq
The linear equations that govern $\psi_{WT}$ are written down in (36) and (37).
The latter implies that
\beq
p + g y + \partial_b \psi_F = 0 \per
\eeq
As $p=0$ after the forcing is turned off, the forced response requires the fluid to change the mean water level $y(b=0)$. If we consider this change to be impulsive, we have
\beq
\frac{\dd y}{\dd t}=\frac{y(\mathcal{T})-y(0)}{\mathcal{T}} = \frac{y}{\mathcal{T}}=-\frac{1}{g\mathcal{T}} \partial_b \psi_F \com
\eeq
\textcolor{black}{where $\mathcal{T}$ is the time scale over which the packet has propagated away. We can explicitly compute $y$ from }
\beq
\left . \partial_b \psi_F \right |_{b=0}
    =  \gamma \int_{-\infty}^{\infty} \frac{\mathcal{A}^2}{(a-a')^2}\dd a'
    = -\gamma \int_{-\infty}^{\infty} \frac{(\mathcal{A}^2)_{a'}}{(a-a')}\dd a' \com
\eeq
\textcolor{black}{where}
\beq
\textcolor{black}{\gamma = \frac{4\omega^3 T^2k^2}{\pi g}}.
\eeq
\textcolor{black}{We recognize the final integral as the Hilbert transform of $\mathcal{A}^2_{a'}$, which we write as $H(\mathcal{A}^2_{a}) $. This change in mean water level sets up a circulation that may be found by linearized equations of motion. To this end, we solve}
\beq
\textcolor{black}{x_{\tau a}-y_{\tau b}=0; \quad x_{\tau b}+y_{\tau a}=0,}
\eeq
\textcolor{black}{and define a final stream function $\psi_4$, so that $\dot{x}=-\psi_{4b}, y_{\tau}=\psi_{4a}$ and }
\beq
\textcolor{black}{\Delta \psi_4 = 0, }
\eeq
\textcolor{black}{and }
\beq
\textcolor{black}{\left. y_{\tau}\right|_{\tau = 0, b=0}=\psi_{4a} 
= \frac{\gamma}{g\mathcal{T}}H(\mathcal{A}^2_a).}
\eeq
\textcolor{black}{$\psi_{4a}$ and $\psi_{4b}$ are now harmonic conjugates, so we may immediately write down their solution with this boundary condition:}
\beq
\textcolor{black}{\psi_{4a} = -\frac{\gamma}{\pi g\mathcal{T}}\int_{-\infty}^{\infty} \frac{\mathcal{A}^2b}{(a-a')^2+b^2}\dd a',}
\eeq
\beq
\textcolor{black}{\psi_{4b} = \frac{\gamma}{\pi g\mathcal{T}}\int_{-\infty}^{\infty} \frac{\mathcal{A}^2 (a-a')}{(a-a')^2+b^2}\dd a'.}
\eeq
\textcolor{black}{We can integrate this to find }
\beq
\textcolor{black}{\psi_4 = -\frac{\gamma}{\pi g \mathcal{T}}\int_{-\infty}^{\infty} \mathcal{A}^2 \tan^{-1}\left(\frac{b}{a-a'}\right)\dd a'+C_0,}
\eeq
\textcolor{black}{where we choose the constant $C_0$ so that $\psi_4(b=0)=0$ implying this flow has no horizontal momentum. }



In figure 3, we show the streamlines when the forcing is turned off and the wave packet propagates away from the generation region. The packet and its return flow have no momentum. In the forcing region, we are left with the dipole that was created during the forcing event. As this can only propagate under its own image (chosen to ensure that the free surface acts as a rigid lid for this flow), this will be extremely weak and the momentum is effectively localized to this forcing region. \\

\begin{figure}
    \centering
    \includegraphics[width = 0.75\textwidth]{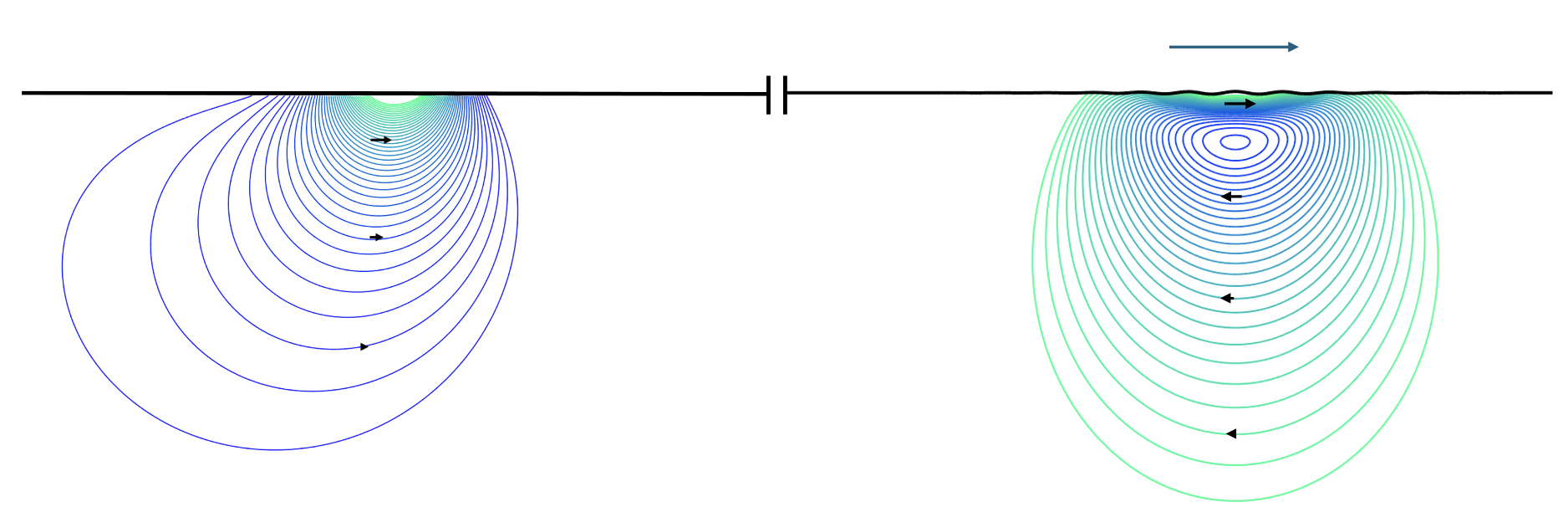}
    \caption{\textcolor{black}{The stream lines after the forcing is turned off. The left shows the streamlines in the forcing region, which consist of the initial long wave generated by the pressure forcing, which is equal and opposite of the long wave for an unforced packet, plus a contribution from a change in mean water level that occurs to satisfy the free surface pressure condition after the forcing is turned off. Right: the streamlines for the waves that have propagated away from the forcing region. The wave packet again has zero net momentum, agreeing with the classical result. In the region of forcing, the momentum is entirely contained in the flow that the pressure disturbance generated that was equal and opposite to the long wave response.}  }
    \label{fig:experiment}
\end{figure}



\subsection{Waves approaching finite depth}

What happens as these waves approach shore? As was shown in \eqref{no_mo}, unforced deep-water wave packets have no momentum. However, \citet{LH1964} showed that if the carrier waves are still in deep-water, but the packet is in finite depth, the wave packet possesses momentum which we denote $I_0$. If a wave packets starts off so that the waves and the packet are in infinitely deep-water and then impinge upon a step such that the depth changes so that the wave packet is now in finite depth, this would seem to violate conservation of momentum, as the packet has experienced a change in momentum, $I_0$. To balance this, somewhere in the fluid there must be momentum with value $-I_0$. As the carrier waves are still in deep-water, the drift (and hence momentum) associated with the carrier waves does not change over the step. Therefore, the long wave is adjusting, and weakening, when it is in finite depth versus deep-water. That excess momentum is jettisoned at the step. \textcolor{black}{As waves do not have momentum in deep-water, this means that a current with momentum $-I_0$ must be induced. This flow can only advect weakly under its own image charge, so that it is essentially bound to the step region.  }

\section{Conclusion}

We find that during the forcing event, the fluid has momentum which is coincident with the Stokes drift of the waves. 
Moreover, the additional momentum that would be associated with the deep return flow is precisely canceled by deep interior pressure gradients sustained by the atmospheric pressure forcing at the surface.
But we also find, like \cite{LH1962}, that the Stokes drift and its associated deep Eulerian return flow are \textit{carried away} from the generation region with the wave packet as soon as the forcing ceases and the packet begins to freely propagate.
This leads to the remarkable conclusion that ultimately, the momentum transferred from the atmosphere --- which is \textit{transmitted} by deep pressure forces during wave generation --- resides in a deep mean flow that identically opposes the deep Eulerian return flow up to the moment that forcing stops. 
In non-rotating, unstratified deep water, this mean flow is non-propagating and thus \textit{remains in place} as the new surface gravity wave packet propagates away. \textcolor{black}{Note, this is invariant to the free surface being a rigid lid and instead is a consequence of the fact that in deep-water irrotational wave packets do not have momentum.}

Besides the modification to the momentum budget considered here, \citet{Dysthe1979} argued that the inclusion of the long wave mean flow modifies the modulation instability growth rates and it was subsequently shown that this gives more accurate predictions that the lower order nonlinear Schrodinger equation \citep{Melville1982, Pizzo2016, Pizzo2019b}.
However, during forcing this long wave is not presented implying a modulation to the instability growth rates. 

A natural question is whether stratification or rotation will significantly change the picture presented here.
If the atmospheric forcing is nearly impulsive, we need consider only the initial value problem in which a surface gravity wave packet is superimposed with a Lagrangian-mean flow given by the Stokes drift, without a deep Eulerian return flow --- the state of the system at the conclusion of an impulsive atmospheric forcing resonant with a surface gravity wave packet (this situation was considered in large eddy simulations with rotation by \cite{wagner2021near}).
In a stratified fluid, the resulting evolution of the freely-propagating packet was analyzed by \cite{Haney2017}, who showed that the deep Eulerian return flow is slightly modified by the presence of realistic oceanic stratfication and in addition weakly radiates internal waves as the packet propagates.
The mean flow left behind by the atmospheric forcing is thus accordingly weakly modified by stratification, in the opposite manner, and likely also decomposes into a weak propagating internal response together with a stronger stationary vortical response.
Less clear is the fate of atmospheric momentum and evolution of a freely-propagating packet in a rotating fluid.
Unlike stratification, rotation likely dramatically changes the induced flow surrounding a freely-propagating packet.
For example, \citet{ursell1950theoretical} (see also Pollard, 1970\nocite{pollard1970generation}) showed that the shallow, Stokes-drift-associated part of the wave-induced mean flow cannot persist in a rotating fluid.
As a result, the circulation around a freely-propagating packet in a rotating fluid is likely much weaker.
We may then speculate that the flow left behind by the packet, which contains the momentum transferred from the atmosphere, takes the form of a packet-scale near-inertial wave with an initial state given by the wave packet Stokes drift profile.\\


\textcolor{black}{We dedicate this paper to Sean Haney. Perhaps the only thing Sean loved more than surface waves was debating their finer details with friends, usually after surf sessions. Sean provided many useful discussions on this topic prior to his untimely passing in 2021. Be careful, Sean. We also thank an anonymous referee for useful comments that have improved the manuscript.} We thank Aidan Blaser and T.S. van den Bremer for helpful comments. N.P. was partially supported by NSF OCE-2219752 and 2342714 and by NASA 80NSSC19K1037 (S-MODE) and 80NSSC23K0985 (OVWST). 


\bibliographystyle{jfm}
\bibliography{ref}

\end{document}